\documentclass[aip,reprint]{revtex4-1}
\usepackage{graphicx}
\usepackage{color}
\usepackage{epsfig}

\begin{document}

\title{Electrodynamics of superconducting pnictide superlattices}

\author{A. Perucchi}
\affiliation{INSTM Udr Trieste-ST and Sincrotrone Trieste, Area Science Park, I-34012
Trieste, Italy} 

\author{F. Capitani}
\affiliation{Dipartimento di Fisica, Universit\`a di Roma Sapienza, Piazzale
Aldo Moro 2, I-00185 Rome, Italy} 

\author{P. Di Pietro}
\affiliation{INSTM Udr Trieste-ST and Sincrotrone Trieste, Area Science Park, I-34012
Trieste, Italy} 

\author{S. Lupi}
\affiliation{CNR-IOM  and Dipartimento di Fisica, Universit\`a di Roma
Sapienza,  P.le Aldo Moro 2, I-00185 Roma, Italy}

\author{S. Lee} 
\author{J.H. Kang}
\affiliation{Department of Materials Science and Engineering, University of 
Wisconsin-Madison, Madison, WI 53706, USA}

\author{J. Jiang}
\author{J.D. Weiss}
\author{E.E. Hellstrom}
\affiliation{Applied Superconductivity Center, National High Magnetic Field 
Laboratory, Florida State University, 2031 East Paul Dirac Drive, 
Tallahassee, FL 32310, USA}

\author{C.B. Eom}
\affiliation{Department of Materials Science and Engineering, University of 
Wisconsin-Madison, Madison, WI 53706, USA}

\author{P. Dore}
\affiliation{CNR-SPIN  and Dipartimento di Fisica, Universit\`a di Roma
Sapienza,  P.le Aldo Moro 2, I-00185 Roma, Italy}

\date{\today}

\date{\today}%

\begin{abstract}
It has been recently reported
(S. Lee {\it et al.}, Nature Materials {\bf 12}, 392, 2013) that superlattices
where layers of the 8\% Co-doped BaFe$_2$As$_2$ superconducting pnictide are 
intercalated with non superconducting ultrathin layers of either SrTiO$_3$  
or of oxygen-rich BaFe$_2$As$_2$, can be used to control flux pinning,
thereby increasing critical fields and currents, without significantly affecting the critical temperature of the pristine superconducting material. 
However, little is known about the electron properties of these systems. Here we investigate the electrodynamics of these superconducting pnictide superlattices 
in the normal and superconducting state by using infrared reflectivity, 
from THz to visible range.
We find that multi-gap structure of these 
superlattices is preserved, whereas some significant changes are observed in their electronic structure with respect to those of the original pnictide. 
Our results suggest that possible attempts to 
further increase the flux pinning may lead to a breakdown of the
pnictide superconducting properties.

\end{abstract}

\maketitle

Iron-based superconductors, as well as high temperature cuprate 
superconductors, are intrinsically multi-layered materials. Particular 
efforts have thus been devoted to the deposition of thin superconducting 
films and to artificially synthesize heterostructures based onto 
different superconducting materials. 
The study of these systems can indeed help in understanding the mechanism 
of superconductivity itself and can offer the possibility of  
tailoring important superconducting properties. 
In particular, high-field applications, which probably represent today 
the most important technological advantage offered by novel
superconductors, require extremely high values of critical 
current J$_c$ and upper critical field H$_c$. While, till now, the enhancement 
of these quantities has been studied in multilayers 
mainly based on the high-T$_c$ cuprate 
YBaCuO,\cite{triscone,haugan} 
only recently epytaxial growth techniques provided artificially engineered 
superlattices (SL) based on the BaFe$_2$As$_2$ (Ba-122) pnictide, 
which enable strong vortex pinning 
and can thus produce very large J$_c$ and H$_c$ enhancements.\cite{lee13} 
In particular, it has been shown that intercalating the Co-doped 
Ba-122 pnictide with non superconducting layers of 
either SrTiO$_3$ (STO-SL) or oxygen-rich undoped 
Ba-122 (O-Ba122-SL), substantially 
enhances the critical currents, especially for magnetic fields parallel 
to the $ab$-plane. On the other hand, the critical temperature does not 
vary significantly, ranging from 19.7 K in STO-SL, 
to 21.9 K in the pristine Co-doped film, 
up to 22.6 K in O-Ba122-SL.\cite{lee13} 

In spite of their interest, little is known about the electronic 
properties of these superlattice (SL) films. 
In particular, it is unclear to what extent 
the multilayered structure affects the multiband character of the Ba-122  
superconductor.\cite{richard,hirschfeld} 
Understanding this point is crucial for the possibility 
of further enhancing the critical fields in pnictide systems, without 
modifying significantly the superconducting properties of the 
pristine material. 

We employ here infrared spectroscopy to study the electronic 
structure of pnictide SLs in the normal and superconducting 
states. Infrared spectroscopy is indeed a powerful technique  
to investigate both charge dynamics and band structure as it 
probes both free carriers and interband excitations and can 
provide a separation of the different contributions to the 
frequency dependent conductivity.\cite{wooten, dressel} 
Furthermore, mesurements performed in 
the THz region can provide direct information about presence and nature
 of superconducting energy gaps.\cite{tinkham,dressel} 
In the case of the Ba-122 system, several infrared studies
have been performed, in particular in the THz region 
(for a recent review, see Ref. [\citenum{perucchi13}])

The three samples under investigation are a 400 nm thick film 
of 8\% Co-doped Ba122 (Co-Ba122-film), 
and the STO-SL and O-Ba122-SL introduced above. 
Both STO-SL and O-Ba122-SL are made of 24 repetitions of a 13 nm 
Co-doped Ba122 layer, intercalated with either 1.3 nm STO or 
3 nm O-Ba122 layer. 
All samples are grown on a substrate, made of a SrTiO$_3$ (STO) template 
(40 nm thick) deposited on (La,Sr)(Al,Ti)O$_3$ (LSAT).   
More details on sample preparation and 
morphology can be found in Ref. [\citenum{lee13}]. 
For comparison THz measurements were also performed on a film 
of undoped Ba-122.

The temperature dependent absolute reflectivities $R(\omega)$ in the 
normal state (25-300 K) are measured with the help of a BRUKER 70v 
interferometer. By using gold and aluminum mirrors as reference and 
various beamsplitters, detectors and thermal sources, $R(\omega)$ spectra  
extend over a frequency range from THz to visible.
To measure the reflectivity in the superconducting state we use THz 
synchrotron radiation from the SISSI beamline \cite{lupi07}
of the Elettra synchrotron. 
The reflectivity ratio $R_S/R_N$ (i.e. the ratio between the 
6 K superconducting and the 25 K normal state reflectivity) is measured  by cycling 
the temperature above and below $T_c$, without moving the sample. 
The measure of the reflectance ratio, as well as that of the transmittance ratio, has been introduced long ago (see the seminal work of Ref. [\citenum{palmer68}] and references therein), 
and is still widely used today in the study of different 
superconducting systems.\cite{perucv3si,perucchi13,xi,xi2} Indeed, by exploiting sinchrotron radiation, this technique can provide high accuracy results intrinsically unaffected by possible misalignments between sample and reference. 

\begin{figure}
\includegraphics[width=8.8 cm]{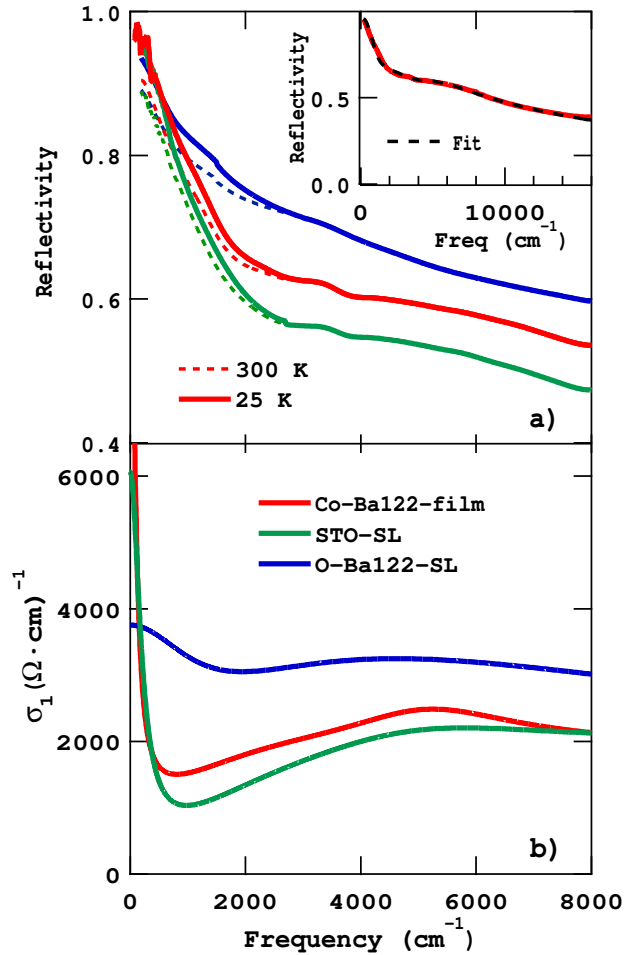}
\caption{Color online. a) Optical Reflectivity $R(\omega$) of the 
three samples (Co-Ba122-film, STO-SL, and O-Ba122-SL) at 25 and 300 K. 
Inset: Full range $R(\omega$) of the Co-doped film, and corresponding 
best fit curve (see text). b) Optical conductivity of the three samples at 25 K, as 
recalculated from the two-band Drude-Lorentz model.}
\label{fig1}
\end{figure}

The reflectivies of the three samples measured at 300 and 25 K 
are reported in Fig. \ref{fig1}a up to 8000 cm$^{-1}$. 
In the inset, the $R(\omega)$ spectrum of the Co-Ba122-film
is shown in all the explored spectral range. 
For all samples, the effect of temperature is only evident 
below 3000 cm$^{-1}$, where the low temperature reflectivities 
are slightly enhanced with respect to room temperature, as a 
typical consequence of a decreased scattering rate.
The $R(\omega)$ of the Co-Ba122-film displays a pronounced edge 
at about 2000 cm$^{-1}$, typically attributed to the multiband 
nature of this material. 
The $R(\omega)$ of the STO-SL is very similar to that of the film, 
though slightly depressed. On the other hand the $R(\omega)$ of the 
O-Ba122-SL has a different shape, with a less pronounced bend at
2000 cm$^{-1}$ and a higher level in the mid- and near-infrared range. 

To describe the complex conductivity of the system, which determines 
its optical response,\cite{wooten,dressel} we use a two-band 
model. While density functional theory \cite{hirschfeld11} predicts 
up to 5 bands crossing the Fermi level, it has been proven that simplified 
models in which only two bands are considered, capture the essential 
low-energy physics in iron-based superconductors.\cite{raghu08,charnukha11}
The model includes the sum of two Drude terms 
(parallel conductivity model \cite{ortolani08})
and of Lorentz terms accounting for possible interband transitions.  
This two-band Drude-Lorentz model has been widely employed 
in describing the optical response of Ba122 system
(see for example Ref. [\onlinecite{perucchi13}] and refs. therein).
In the present case, the fitting procedure\cite{footnote} takes into account, 
besides the finite thickness of the samples, the possible 
reflections from the substrate via Fresnel equations.\cite{dressel} 
This procedure requires the knowledge of the optical properties of 
the substrate, which were obtained from \textit{ad-hoc} measurements 
on the STO+LSAT substrate by using a standard Lorentz-model 
fitting.\cite{wooten,dressel}  

The $R(\omega)$ spectra can be fitted by using, besides the two 
Drude terms, a first broad Lorentz term mimicking electronic 
interband transitions in both mid- and near-infrared, and a second 
one accounting for contributions coming from 
the ultraviolet.\cite{lucarelli10,nakajima10} An example of the best fit curve is reported in the inset 
of Fig. \ref{fig1}. 
 The optical conductivity $\sigma_1(\omega)$ 
 is recalculated according to the model and displayed in Fig. \ref{fig1}b. 
A clear separation between the two Drude bands is well distinguishable for both the Co-Ba122-film and the STO-SL, since the  
scattering rates of these two Drude bands are about 100$\div$200 cm$^{-1}$ 
(narrow Drude) and 4000 cm$^{-1}$ (broad Drude).
We recall that the presence of two Drude bands with very different
widths has been adopted  by many authors to describe the optical
properties of BaFe$_{2-x}$Co$_{x}$As$_2$  systems 
(see for example Refs. [\onlinecite{lucarelli10,nakajima10,baldassarre12}]).
The broad peak centered at 5000 cm$^{-1}$ can be associated
to an electronic interband transition.\cite{nakajima10} 
In the O-Ba122-SL case, the situation is much different.  
The first Drude term becomes strongly damped and less discernible from 
the electronic background due to the presence of the second broad Drude 
component. A brodening of the peak at 5000 cm$^{-1}$ is also observed.

The increased scattering rate observed in the normal state of the 
O-Ba122-SL  seems to be at odds with the better crystalline quality of 
this compound with respect to STO-SL, as revealed by x-rays.
As reported in Ref. [\citenum{lee13}], the crystalline quality is influenced 
by the better structural matching of the O-Ba122 layers on the Co-Ba122 
matrix. To reconcile this scenario with our observations, we note that not 
only the narrow Drude band scattering rate increases, but also the overall 
optical conductivity is strongly enhanced in the O-Ba122-SL. This should 
be interpreted as an increase in the number of charge carriers probably 
due to the oxygen enrichment, acting like a dopant. 
Moreover, the broadening of the peak at 
5000 cm$^{-1}$ also indicates non-negligible effects on the band structure. 
Qualitatively, we may think that while the band structure of STO and Co-Ba122 
are fully disconnected thus creating energetic barriers between the various 
portions of the sample, the situation with O-Ba122 is much more different. 
In fact one may expect the bands to be continuously connected (band bending), 
with a consequent modification of both band structure and chemical potential 
in the proximity of the interface. 
The presence of the O-Ba122 intercalation thus seems to affect the electronic 
structure in a more profound way, beyond the possible issues related 
to crystallinity and disorder. 

\begin{figure}
\includegraphics[width=8.8 cm]{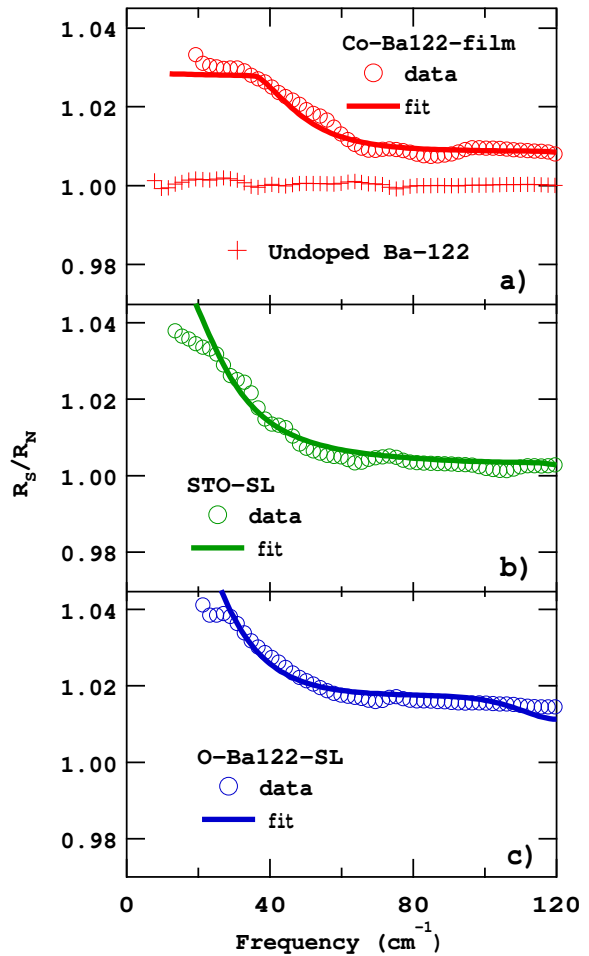}
\caption{Color online. a) Reflectivity ratio $R_S/R_N = R(6 K)/R(25 K)$ for an undoped and a 8\%Co-doped Ba122 film (Co-Ba122-film). 
The full red line is the best fit curve of the Co-Ba122-film data (see text). 
b) Reflectivity ratio and best fit curve for the STO-SL. 
c) Same as before for the O-Ba122-SL.}
\label{fig2}
\end{figure}

One may thus wonder whether multiband superconductivity is preserved at 
all in the superlattice compounds and in particular in O-Ba122-SL. 
To answer this question we turn to the THz electrodynamics in the 
superconducting state. We compare in Fig. \ref{fig2}a the measured 
reflectivity ratio $R_S/R_N$ for the Co-Ba122-film, together with that of the undoped, non superconducting, Ba122 film. 
As expected,\cite{tinkham,dressel} the $R_S/R_N$ spectrum clearly deviates 
from a straight line, as a consequence of superconducting gap(s) opening
in the Co-Ba122-film case, while it is flat and very close to 1 in the case 
of the non-superconducting Ba122 film.
It is worth noting that the $R_S/R_N$ spectral shape is very similar to those reported in previous works on different Co-doped Ba122 films.\cite{perucchi10,perucchi13} 

To describe the electrodynamics of the superconducting systems here investigated, the same procedure employed in the analysis of the THz response of different two-band systems (such as Co-doped Ba122 films,\cite{perucchi10,perucchi13} MgB$_2$,\cite{ortolani08} and V$_3$Si \cite{perucv3si}) has been used.
In this procedure, each Drude term is substituted by the Zimmermann term,\cite{zimmermann91} which describes the electrodynamics of a superconductor of arbitrary purity, thus generalizing the 
standard BCS Mattis-Bardeen model.\cite{tinkham,dressel} 
By employing the same Drude-Lorentz 
normal state parameters obtained from the fittings of Fig.\ref{fig1}, 
we obtain superconducting gap values $\Delta_A$=18 and 
$\Delta_B$= 60 cm$^{-1}$, for the narrow and broad Drude terms, respectively.

In the case of the STO-SL (Fig.\ref{fig2}b) and O-Ba122-SL (Fig.\ref{fig2}c), 
$R_S/R_N$ has a shape which is reminding that of the Co-Ba122-film. 
The main difference is in the onset of the $R_S/R_N$ enhancement, 
which is found at about 40 cm$^{-1}$, slightly lower than in the Co-Ba122-film 
($\approx$60 cm$^{-1}$). The shape of $R_S/R_N$ at the lowest frequencies 
(i.e. below 25 cm$^{-1}$), is more flat in the Co-Ba122-film, with respect 
to the two SLs.
The fitting results \cite{footnote} (see 
Figs. \ref{fig2}b and c)
 show that, in both cases, the smaller gap $\Delta_A$ reduces down to about  to 7 cm$^{-1}$, while $\Delta_B$ essentially remains unchanged. 
 
 It is clear that the inclusion in the model of an additional third 
electronic channel, ungapped or characterized by a very small 
gap, may further increase the fit quality, especially on the low frequency side. 
However, the present model already captures the essential physics we are 
interested in. In particular, the main message is that the multi-gap
structure in the superconducting state is preserved in both SLs. 
This result is far from trivial in the light of the important differences
observed in the normal state properties, especially for what concerns 
the O-Ba122-SL. The main difference between the Co-doped Ba122 film and the 
two SLs is the decrease of the smaller gap value $\Delta_A$. 
Interestingly, such a decrease does not affect significantly neither 
the larger gap $\Delta_B$ nor $T_C$. 

We have addressed here the electrodynamic properties of the pnictide 
superlattices STO-SL and O-Ba122-SL. 
The infrared data show that the normal state electronic properties of the STO-SL are very similar to those of the Co-Ba122-film. On the contrary, the O-Ba122-SL has different electronic properties with, in particular, a 
less pronounced multi-band structure. However, we have verified that both SL
systems remain multi-gap superconductors, even though the small 
gap value $\Delta_A$ is slightly reduced with respect to the value observed in the Co-Ba122-film. In conclusion, our results point out that, albeit the 
multi-gap structure is still preserved in both SLs, one should be 
aware of the risks related to changes occurring in the electronic properties 
while attempting to further increase pinning centers, especially in 
the case of the oxygen enrichment technique.

\begin{acknowledgments}

This work was partially supported by Italian Ministry of Research (MIUR) program FIRB Futuro in Ricerca grant
no. RBFR10PSK4. Work at the University of Wisconsin was financially supported by the DOE Office of Basic
Energy Sciences under award number DE-FG02-06ER46327. Work at the Sapienza University of Rome was partially supported  by the MIUR PRIN2012 Project No.2012X3YFZ2.

\end{acknowledgments}

\end{document}